\documentclass[prl,article,twocolumn,reprint,superscriptaddress]{revtex4}
\usepackage{graphicx,amsfonts,color,comment,amsmath,hyperref,float}
\usepackage{dcolumn}
\usepackage{graphicx}
\usepackage{natbib,aas_macros}
\usepackage{color}
\usepackage{hyperref}

\newcommand{\msun}{M_{\odot}}

\begin{document}
\title{Hidden Cores of Active Galactic Nuclei as the Origin of Medium-Energy Neutrinos:\\
Critical Tests with the MeV Gamma-Ray Connection}

\author{Kohta Murase}
\affiliation{Department of Physics, The Pennsylvania State University, University Park, Pennsylvania 16802, USA}
\affiliation{Department of Astronomy \& Astrophysics, The Pennsylvania State University, University Park, Pennsylvania 16802, USA}
\affiliation{Center for Multimessenger Astrophysics, Institute for Gravitation and the Cosmos, The Pennsylvania State University, University Park, Pennsylvania 16802, USA}
\affiliation{Yukawa Institute for Theoretical Physics, Kyoto, Kyoto 606-8502 Japan}
\author{Shigeo S. Kimura}
\affiliation{Department of Physics, The Pennsylvania State University, University Park, Pennsylvania 16802, USA}
\affiliation{Department of Astronomy \& Astrophysics, The Pennsylvania State University, University Park, Pennsylvania 16802, USA}
\affiliation{Center for Multimessenger Astrophysics, Institute for Gravitation and the Cosmos, The Pennsylvania State University, University Park, Pennsylvania 16802, USA}
\author{Peter M\'esz\'aros}
\affiliation{Department of Physics, The Pennsylvania State University, University Park, Pennsylvania 16802, USA}
\affiliation{Department of Astronomy \& Astrophysics, The Pennsylvania State University, University Park, Pennsylvania 16802, USA}
\affiliation{Center for Multimessenger Astrophysics, Institute for Gravitation and the Cosmos, The Pennsylvania State University, University Park, Pennsylvania 16802, USA}

\date{submitted 8 April 2019; accepted 26 May 2020}

\begin{abstract}
Mysteries about the origin of high-energy cosmic neutrinos have deepened by the recent IceCube measurement of a large diffuse flux in the 10--100~TeV range. Based on the standard disk-corona picture of active galactic nuclei (AGN), we present a phenomenological model enabling us to systematically calculate the spectral sequence of multimessenger emission from the AGN coronae. We show that protons in the coronal plasma can be stochastically accelerated up to PeV energies by plasma turbulence, and find that the model explains the large diffuse flux of medium-energy neutrinos if the cosmic rays carry only a few percent of the thermal energy. We find that the Bethe-Heitler process plays a crucial role in connecting these neutrinos and cascaded MeV gamma rays, and point out that the gamma-ray flux can even be enhanced by the reacceleration of secondary pairs. Critical tests of the model are given by its prediction that a significant fraction of the MeV gamma-ray background correlates with $\sim10$~TeV neutrinos, and nearby Seyfert galaxies including NGC 1068 are promising targets for IceCube, KM3Net, IceCube-Gen2, and future MeV gamma-ray telescopes.
\end{abstract}

\maketitle

%
The origin of cosmic neutrinos observed in IceCube is a major enigma~\cite{Aartsen:2013bka,Aartsen:2013jdh}, and the latest data of high- and medium-energy starting events and shower events~\cite{Aartsen:2014muf,Aartsen:2015ita,Aartsen:2020aqd} are more puzzling. The atmospheric background of high-energy electron neutrinos is lower than that of muon neutrinos, allowing us to analyze the data below 100~TeV~\cite{Beacom:2004jb,Laha:2013lka}. The extragalactic neutrino background (ENB) at these energies has shown a larger flux with a softer spectrum, compared to the $\gtrsim100$~TeV data~\cite{Aartsen:2016xlq,ICRCtrack}.
The comparison with the extragalactic gamma-ray background (EGB) measured by {\it Fermi} indicates that the 10--100~TeV ENB originates from {\it hidden} sources preventing the escape of GeV--TeV gamma rays~\cite{Murase:2015xka}. 

Active galactic nuclei (AGN) are major contributors to the energetics of high-energy cosmic radiations~\cite{Murase:2018utn}; radio quiet (RQ) AGN are dominant in the extragalactic x-ray sky~\cite{1992ARA&A..30..429F,Ueda:2003yx,Hasinger:2005sb,Ajello:2008xb,Ueda:2014tma}, and jetted AGN that are typically radio loud (RL) dominantly explain the EGB~\cite{Costamante:2013sva,Inoue:2014ona,Fornasa:2015qua}. 
AGN may also explain the MeV gamma-ray background whose origin has been under debate (e.g., Refs.~\cite{Inoue:2007tn,Ajello:2009ip,Lien:2012gz}).

High-energy neutrino production in the vicinity of supermassive black holes (SMBHs) were discussed early on~\cite{Eichler:1979yy,1981MNRAS.194....3B,brs90,Stecker:1991vm}, in particular to explain x-ray emission by cosmic-ray (CR) induced cascades assuming the existence of high Mach number accretion shocks at the inner edge of the disk~\cite{1986ApJ...304..178K,1986ApJ...305...45Z,1987ApJ...320L..81S,Stecker:1991vm}. However, cutoff features evident in the x-ray spectra of Seyfert galaxies and the absence of electron-positron annihilation lines ruled out the simple cascade scenario for the x-ray origin (e.g., Refs.~\cite{1995ApJ...438..672M,2018MNRAS.480.1819R}). In the standard disk-corona scenario, the observed x rays are attributed to thermal Comptonization of disk photons~\cite{Shapiro:1976fr,1980A&A....86..121S,Zdziarski:1996wq,1996ApJ...470..249P,Haardt:1996hn}, and electrons are presumably heated in the coronal region~\cite{1977ApJ...218..247L,Galeev:1979td}. There has been significant progress in our understanding of accretion disks with the identification of the magnetorotational instability (MRI)~\cite{bh91,BH98a}, which can result in the formation of a corona above the disk as a direct consequence of the accretion dynamics and magnetic dissipation (e.g., Refs.~\cite{Miller:1999ix,Merloni:2000gs,Liu:2002ts,Blackman:2009fi,Io:2013gja,SI14a,Jiang:2014wga}).  

Accompanied turbulence and magnetic reconnections are important for particle acceleration~\cite{Lazarian:2012nd}. The roles of nonthermal particles have been studied in the context of radiatively inefficient accretion flows (RIAFs)~\cite{ny94,YN14a}, in which the plasma is often collisionless because Coulomb collisions are negligible for protons (e.g., Refs.~\cite{tk85,Mahadevan:1997qz,Mahadevan:1997zq,ktt14,lyn+14,Ball:2017bpa}). Recent studies based on numerical simulations of the MRI~\cite{Kimura:2016fjx,Kimura:2018clk} support the idea that high-energy ions may be accelerated in the presence of the magnetohydrodynamic (MHD) turbulence.

\begin{figure}
\begin{center}
\includegraphics[width=0.9\linewidth]{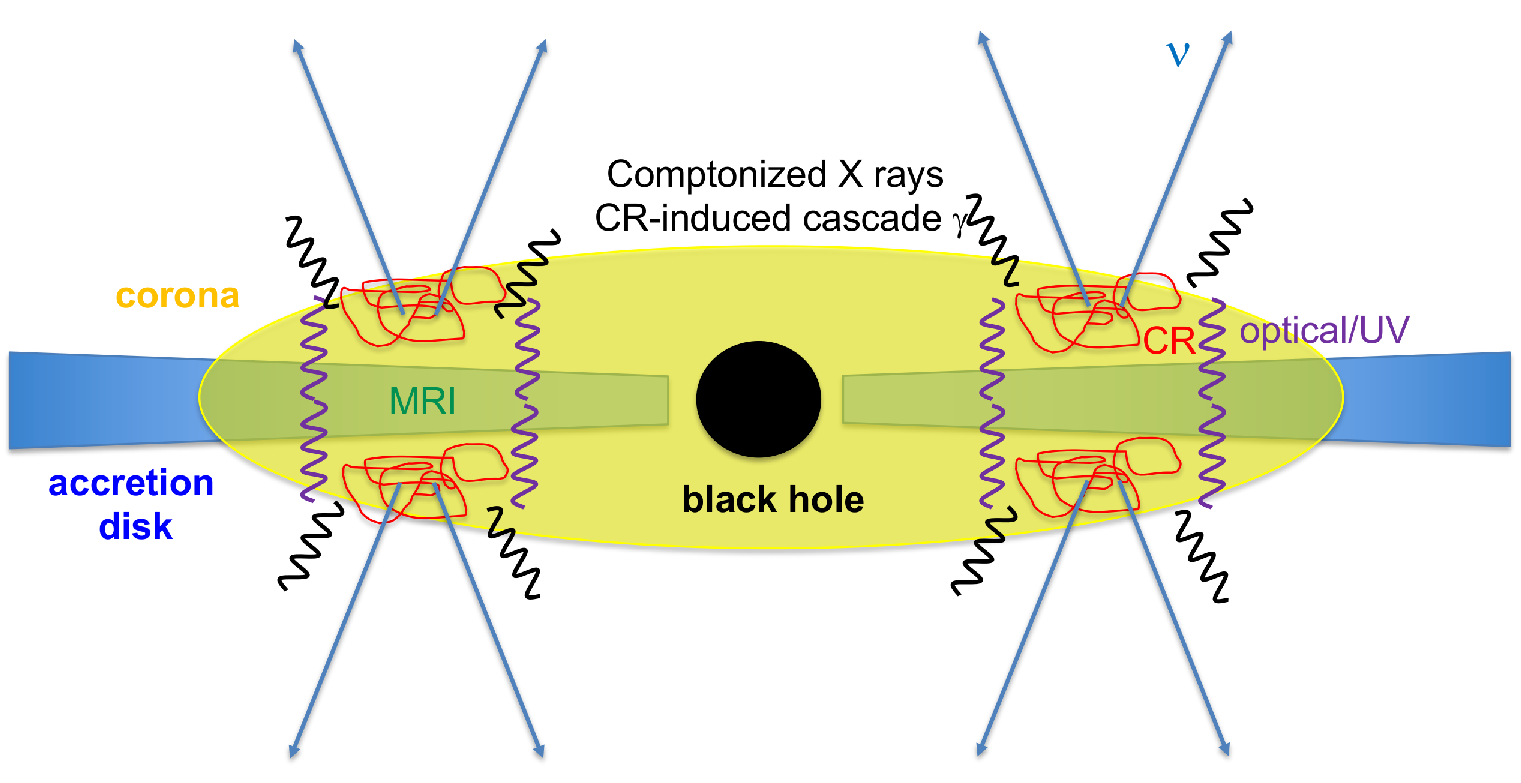}
\caption{Schematic picture of the AGN disk-corona scenario. Protons are accelerated by plasma turbulence generated in the coronae, and produce high-energy neutrinos and cascaded gamma rays via interactions with matter and radiation.}
\label{fig:model}
\end{center}
\end{figure}

The vicinity of SMBHs is often optically thick to GeV--TeV gamma rays, so that CR acceleration~\footnote{In this work CRs are used for nonthermal ions and nucleons that do not have to be observed on Earth.} cannot be directly probed by these photons, but high-energy neutrinos can be used as a unique probe of the physics of AGN cores. In this work, we present a new concrete model for these high-energy emissions (see Fig.~\ref{fig:model}). Spectral energy distributions (SEDs) are constructed from the data and from empirical relations, and then we compute neutrino and cascade gamma-ray spectra by consistently solving particle transport equations. 
We demonstrate the importance of future MeV gamma-ray observations for revealing the origin of IceCube neutrinos especially in the medium-energy ($\sim10-100$~TeV) range and for testing neutrino emission from NGC 1068 and other AGN.

We use a notation with $Q_x=Q\times10^x$ in CGS units.

%
{\it Phenomenological prescription of AGN disk coronae.---}We begin by providing a phenomenological disk-corona model based on the existing data. 
Multiwavelength SEDs of Seyfert galaxies have been extensively studied, consisting of several components; radio emission (see Ref.~\cite{2019arXiv190205917P}), infrared emission from a dust torus~\cite{Netzer:2015jna}, optical and ultraviolet components from an accretion disk~\cite{1999PASP..111....1K}, and x rays from a corona~\cite{1980A&A....86..121S}. 
The latter two components are relevant for this work.

The ``blue'' bump, which has been seen in many AGN, is attributed to multitemperature blackbody emission from a geometrically thin, optically thick disk~\cite{Shakura:1972te}.  
The averaged SEDs are provided in Ref.~\cite{ho08} as a function of the Eddington ratio, $\lambda_{\rm Edd}=L_{\rm bol}/L_{\rm Edd}$, where $L_{\rm bol}$ and $L_{\rm Edd}\approx1.26\times{10}^{45}~{\rm erg}~{\rm s}^{-1}{(M/10^7M_\odot)}$ are bolometric and Eddington luminosities, respectively, and $M$ is the SMBH mass. The disk component is expected to have a cutoff in the ultraviolet range. 
Hot thermal electrons in a corona, with an electron temperature of $T_e\sim10^9$~K, energize the disk photons by Compton upscattering. The consequent x-ray spectrum can be described by a power law with an exponential cutoff, in which the photon index ($\Gamma_X$) and the cutoff energy ($\varepsilon_{X, \rm cut}$) can also be estimated from $\lambda_{\rm Edd}$~\cite{2017MNRAS.470..800T,2018MNRAS.480.1819R}. Observations have revealed the relationship between the x-ray luminosity $L_X$ and $L_{\rm bol}$~\cite{2007ApJ...654..731H} [where one typically sees $L_X\sim(0.01-0.1) L_{\rm bol}$], by which the disk-corona SEDs can be modeled as a function of $L_{X}$ and $M$. 
In this work, we consider contributions from AGN with the typical SMBH mass for a given $L_X$, using $M\approx2.0\times10^7~M_\odot~(L_X/1.16\times10^{43}~{\rm erg}~{\rm s}^{-1})^{0.746}$~\cite{2018arXiv180306891M}. 
The resulting disk-corona SED templates in our model are shown in Fig.~\ref{fig:sed} (see Supplemental Material for details), which enables us to quantitatively evaluate CR, neutrino and cascade gamma-ray emission. 

Next we estimate the nucleon density $n_p$ and coronal magnetic field strength $B$.  
Let us consider a corona with the radius $R\equiv{\mathcal R}R_S$ and the scale height $H$, where ${\mathcal R}$ is the normalized coronal radius and $R_{S}=2GM/c^2$ is the Schwarzschild radius. 
Then the nucleon density is expressed by $n_p\approx\tau_T/(\sigma_TH)$, where $\tau_T$ is the Thomson optical depth that is typically $\sim0.1-1$. The standard accretion theory~\cite{pri81,KFM08a} gives the coronal scale height $H\approx(C_s/V_K){\mathcal R}R_S={\mathcal R}R_S/\sqrt{3}$, where $C_s=\sqrt{k_BT_p/m_p}=c/\sqrt{6\mathcal R}$ is the sound velocity, and $V_K=\sqrt{GM/R}=c/\sqrt{2\mathcal R}$ is the Keplerian velocity. 
For an optically thin corona, the electron temperature is estimated by $T_e\approx\varepsilon_{X,\rm cut}/(2k_B)$, and $\tau_T$ is empirically determined from $\Gamma_X$ and $k_BT_e$~\cite{2018MNRAS.480.1819R}. 
We expect that thermal protons are at the virial temperature $T_p={GMm_p/(3{\mathcal R}R_Sk_B)}=m_pc^2/(6{\mathcal R}k_B)$, implying that the corona may be characterized by two temperatures, i.e., $T_p>T_e$~\cite{DiMatteo:1997cy,Cao:2008by}. 
Finally, the magnetic field is given by $B=\sqrt{8\pi n_pk_BT_p/\beta}$ with plasma beta ($\beta$). 

Many physical quantities (including the SEDs) can be estimated observationally and empirically. Thus, for a given $L_X$, parameters characterizing the corona (${\mathcal R}$, $\beta$, $\alpha$) are remaining. They are also constrained in a certain range by observations~\cite{2012MNRAS.420.1825J,Morgan:2012df} and numerical simulations~\cite{Io:2013gja,Jiang:2014wga}. For example, recent MHD simulations show that $\beta$ in the coronae can be as low as 0.1--10 (e.g., Refs.~\cite{Miller:1999ix,SI14a}). We assume $\beta\lesssim1-3$ and $\alpha=0.1$ for the viscosity parameter~\cite{Shakura:1972te}, and adopt ${\mathcal R}=30$.

\begin{figure}[tb]
\begin{center}
\includegraphics[width=0.85\linewidth]{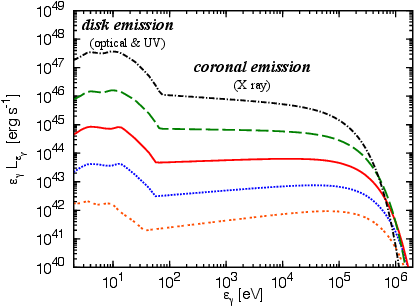}
\caption{Disk-corona SEDs used in this work, for $L_X=10^{42}$, $10^{43}$, $10^{44}$, $10^{45}$, and $10^{46}\rm~erg~s^{-1}$ (from bottom to top). See text for details.}
\label{fig:sed}
\end{center}
\end{figure}

%
{\it Stochastic proton acceleration in coronae.---}Standard AGN coronae are magnetized and turbulent, in which it is natural that protons are stochastically accelerated via plasma turbulence or magnetic reconnections. In this work, we solve the known Fokker-Planck equation that can describe the second order Fermi acceleration process (e.g., Refs.~\cite{Becker:2006nz,Stawarz:2008sp,cc70,1996ApJS..103..255P}). 
Here we describe key points in the calculations of CR spectra (see Supplemental Material or an accompanying paper~\cite{Kimura:2019yjo} for technical details). The stochastic acceleration time is given by $t_{\rm acc}\approx \eta{(c/V_A)}^2(H/c){(\varepsilon_p/eBH)}^{2-q}$, where $V_A$ is the Alfv\'en velocity and $\eta$ is the inverse of the turbulence strength~\cite{Dermer:1995ju,Dermer:2014vaa}. 
We consider $q\sim3/2-5/3$, which is not inconsistent with the recent simulations~\cite{Kimura:2018clk}, together with $\eta\sim10$. 
The stochastic acceleration process is typically slower than the first order Fermi acceleration, which competes with cooling and escape processes. 
We find that for luminous AGN the Bethe-Heitler pair production ($p\gamma\rightarrow pe^+e^-$) is the most important cooling process because of copious disk photons, which determines the proton maximum energy. For our model parameters, the CR spectrum has a cutoff at $\varepsilon_p\sim0.1-1$~PeV, leading to a cutoff at $\varepsilon_\nu\sim5-50$~TeV in the neutrino spectrum. Note that all the loss timescales can uniquely be evaluated within our disk-corona model, and this result is only sensitive to $\eta$ and $q$ for a given set of coronal parameters. 
Although the resulting CR spectra (that are known to be hard) are numerically obtained in this work, we stress that spectra of $p\gamma$ neutrinos are independently predicted to be hard, because the photomeson production occurs only for protons whose energies exceed the pion production threshold~\cite{Murase:2015xka,Kimura:2019yjo}. 
The CR pressure to explain the neutrino data turns out to be $\sim(1-10)$\% of the thermal pressure, by which the normalization of CRs is set.  

For coronae considered here, the infall and dissipation times are 
$t_{\rm fall}\simeq2.5\times{10}^{6}~{\rm s}~\alpha_{-1}^{-1}{({\mathcal R}/30)}^{3/2}R_{S,13.5}$ 
and 
$t_{\rm diss}\simeq1.7\times{10}^{5}~{\rm s}~{(\mathcal R/30)}^{3/2}R_{S,13.5}\beta^{1/2}$, 
respectively.
The Coulomb relaxation timescales for protons 
[e.g., $t_{C,pe}\sim4\times{10}^5~{\rm s}~({\mathcal R}/30)R_{S,13.5}{(\tau_T/0.5)}^{-1}{(k_BT_e/0.1~{\rm MeV})}^{3/2}$] are longer than $t_{\rm diss}$ (especially for $\beta\lesssim1$), so turbulent acceleration may operate for protons rather than electrons (and acceleration by small-scale magnetic reconnections may occur~\cite{Hoshino:2015cka,Li:2015bga}).
This justifies our assumption on CR acceleration (cf. Refs.~\cite{Ozel:2000wm,Kimura:2014jba,Ball:2016mjx,Kimura:2019yjo} for RIAFs). 

%
{\it Connection between 10--100~TeV neutrinos and MeV gamma rays.---}
Accelerated CR protons interact with matter and radiation modeled in the previous section, producing secondary particles. 
We compute neutrino and gamma-ray spectra as a function of $L_X$, by utilizing the code to solve kinetic equations with electromagnetic cascades taken into account~\cite{Murase:2017pfe,Murase:2018okz}. Secondary injections by the Bethe-Heitler and $p\gamma$ processes are approximately treated as $\varepsilon_{e}^2 (d\dot{N}_e^{\rm BH}/d\varepsilon_e)|_{\varepsilon_{e}=(m_e/m_p)\varepsilon_p} \approx t_{\rm BH}^{-1}\varepsilon_p^2(dN_{\rm CR}/d\varepsilon_p)$~\cite{1992ApJ...400..181C,SG83a,Murase:2010va}, $\varepsilon_{e}^2(d\dot{N}_e^{p\gamma}/d\varepsilon_e)|_{\varepsilon_{e}=0.05\varepsilon_p} \approx (1/3)\varepsilon_{\nu}^2(d\dot{N}_\nu^{p\gamma}/d\varepsilon_\nu)|_{\varepsilon_{\nu}=0.05\varepsilon_p} \approx(1/8)t_{p\gamma}^{-1} \varepsilon_p^2(dN_{\rm CR}/d\varepsilon_p)$, 
and $\varepsilon_{\gamma}^2(d\dot{N}_\gamma^{p\gamma}/d\varepsilon_\gamma)|_{\varepsilon_{\gamma}=0.1\varepsilon_p} \approx(1/2)t_{p\gamma}^{-1} \varepsilon_p^2(dN_{\rm CR}/d\varepsilon_p)$. 
The cascade photon spectra are broad, being determined by the energy reprocessing via two-photon annihilation, synchrotron radiation, and inverse Compton emission.

The EGB and ENB are numerically calculated via the line-of-sight integral with the convolution of the x-ray luminosity function given by Ref.~\cite{Ueda:2014tma} (see also Supplemental Material, which includes Refs.~\cite{Murase:2014foa,Ajello:2015mfa,Murase:2008yt,Kotera:2009ms,Fang:2017zjf,Inoue:2011bm,Chen:2014gxa,Palladino:2018evm}). Note that the luminosity density of AGN evolves as redshift $z$, with a peak around $z\sim1-2$, and our prescription enables us to {\it simultaneously} predict the x-ray background, EGB and ENB. 
The results are shown in Fig.~\ref{fig:diffuse}, and our AGN corona model can explain the ENB at $\sim30$~TeV energies with a steep spectrum at higher energies (due to different proton maximum energies), possibly simultaneously with the MeV EGB. 
We find that the required CR pressure ($P_{\rm CR}$) is only $\sim1$\% of the thermal pressure ($P_{\rm th}$), so the energetics requirement is not demanding in our AGN corona model (see Supplemental Material).

\begin{figure}[tb]
\begin{center}
\includegraphics[width=\linewidth]{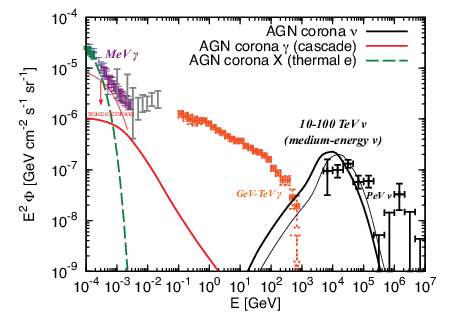}
\caption{EGB and ENB spectra in our AGN corona model. 
The data are taken from {\it Swift} BAT~\cite{Ajello:2008xb} (green), Nagoya balloon~\cite{1975Natur.254..398F} (blue), SMM~\cite{1997AIPC..410.1223W} (purple), COMPTEL~\cite{2000AIPC..510..467W} (gray), {\it Fermi} LAT~\cite{Ackermann:2014usa} (orange), and IceCube shower events (black)~\cite{Aartsen:2020aqd} (consistent with the global fit~\cite{Aartsen:2015ita}). 
Solid thick (thin) curves are for $\beta=1$ and $q=5/3$ ($\beta=3$ and $q=3/2$ with the reacceleration contribution), respectively. 
}
\label{fig:diffuse}
\end{center}
\end{figure}

Remarkably, we find that high-energy neutrinos are produced by {\it both} $pp$ and $p\gamma$ interactions. 
The disk-corona model indicates $\tau_T\approx n_p\sigma_T {\mathcal R}R_S/\sqrt{3}\sim0.1-1$, leading to the effective $pp$ optical depth 
\begin{eqnarray}
f_{pp}&\approx&t_{\rm esc}/t_{pp}\approx n_p (\kappa_{pp}\sigma_{pp})R(c/V_{\rm fall})\nonumber\\
&\sim&2~(\tau_T/0.5)\alpha_{-1}^{-1}{(\mathcal R/30)}^{1/2},
\end{eqnarray}
where $\sigma_{pp}\sim4\times{10}^{-26}~{\rm cm}^2$ is the $pp$ cross section, $\kappa_{pp}\sim0.5$ is the proton inelasticity, and $V_{\rm fall}=\alpha V_K$ is the infall velocity. 
Coronal x rays provide target photons for the photomeson production, whose effective optical depth~\cite{Murase:2008mr,Murase:2015xka} for $\tau_T\lesssim1$ is
\begin{eqnarray}
f_{p \gamma}&\approx&t_{\rm esc}/t_{p\gamma}\approx\eta_{p\gamma}\hat{\sigma}_{p\gamma}R(c/V_{\rm fall}) n_X{(\varepsilon_p/\tilde{\varepsilon}_{p\gamma-X})}^{\Gamma_X-1}\nonumber\\
&\sim&2~\frac{\eta_{p\gamma}L_{X,44}{(\varepsilon_p/\tilde{\varepsilon}_{p\gamma-X})}^{\Gamma_X-1}}{\alpha_{-1}({\mathcal R/30)}^{1/2}R_{S,13.5}(\varepsilon_X/1~{\rm keV})},
\end{eqnarray}
where $\eta_{p\gamma}\approx2/(1+\Gamma_X)$, $\hat{\sigma}_{p\gamma}\sim0.7\times{10}^{-28}~{\rm cm}^2$ is the attenuation $p\gamma$ cross section, $\bar{\varepsilon}_\Delta\sim0.3$~GeV, $\tilde{\varepsilon}_{p\gamma-X}=0.5m_pc^2\bar{\varepsilon}_{\Delta}/\varepsilon_X\simeq0.14~{\rm PeV}~{(\varepsilon_X/1~{\rm keV})}^{-1}$, and $n_X\sim L_X/(2\pi R^2 c \varepsilon_X)$ is used. 
The total meson production optical depth is given by $f_{\rm mes}=f_{p\gamma}+f_{pp}$, which always exceeds unity in our model. 
Note that the spectrum of $p\gamma$ neutrinos should be hard at low energies, because only sufficiently high-energy protons can produce pions via $p\gamma$ interactions with x-ray photons.

Note that $\sim10-100$~TeV neutrinos originate from $\sim0.2-2$~PeV CRs. Unlike in previous studies explaining the IceCube data~\cite{Stecker:2013fxa,Kalashev:2015cma}, here in fact the disk photons are not much relevant for the photomeson production because its threshold energy is $\tilde{\varepsilon}_{p\gamma-\rm th}\simeq3.4~{\rm PeV}~{(\varepsilon_{\rm disk}/10~{\rm eV})}^{-1}$.
Rather, CR protons responsible for the medium-energy neutrinos should efficiently interact via the Bethe-Heitler process because the characteristic energy is $\tilde{\varepsilon}_{\rm BH-disk}\approx0.5m_pc^2\bar{\varepsilon}_{\rm BH}/\varepsilon_{\rm disk}\simeq0.47~{\rm PeV}~{(\varepsilon_{\rm disk}/10~{\rm eV})}^{-1}$, where $\bar{\varepsilon}_{\rm BH}\sim10(2m_ec^2)\sim10$~MeV~\citep{1992ApJ...400..181C,SG83a,Murase:2010va}. 
With the disk photon density $n_{\rm disk}\sim L_{\rm disk}/(2\pi R^2 c \varepsilon_{\rm disk})$ for $\tau_T\lesssim1$, the effective Bethe-Heitler optical depth (with $\hat{\sigma}_{\rm BH}\sim0.8\times{10}^{-30}~{\rm cm}^2$) is
\begin{eqnarray}
f_{\rm BH}&\approx&n_{\rm disk}\hat{\sigma}_{\rm BH}R(c/V_{\rm fall})\nonumber\\
&\sim&40~L_{\rm disk,45.3}\alpha_{-1}^{-1}{(\mathcal R/30)}^{-1/2}R_{S,13.5}^{-1}{(10~{\rm eV}/\varepsilon_{\rm disk})},\,\,\,\,\, 
\end{eqnarray}
which is much larger than $f_{p\gamma}$. 
The dominance of the Bethe-Heitler cooling is a direct consequence of the observed disk-corona SEDs. 
The 10--100~TeV neutrino flux is suppressed by $\sim f_{\rm mes}/f_{\rm BH}$, predicting the tight relationship with the MeV gamma-ray flux.
 
Analytically, the medium-energy ENB flux is given by
\begin{eqnarray}
E_\nu^2\Phi_\nu&\sim&{10}^{-7}~{\rm GeV}~{\rm cm}^{-2}~{\rm s}^{-1}~{\rm sr}^{-1}~\left(\frac{2K}{1+K}\right)\mathcal{R}_p^{-1}\left(\frac{\xi_z}{3}\right)\nonumber\\
&\times&\left(\frac{15f_{\rm mes}}{1+f_{\rm BH}+f_{\rm mes}}\right){\left(\frac{\xi_{\rm CR,-1}L_X\rho_X}{2\times{10}^{46}~{\rm erg}~{\rm Mpc}^{-3}~{\rm yr}^{-1}}\right)}.\,\,\,\,\,\,\,\,
\label{eq:diffuse}
\end{eqnarray}
which is indeed consistent with the numerical results shown in Fig.~\ref{fig:diffuse}.
Here $K=1$ and $K=2$ for $p\gamma$ and $pp$ interactions, respectively, $\xi_z\sim3$ due to the redshift evolution of the AGN luminosity density~\cite{Waxman:1998yy,Murase:2016gly}, ${\mathcal R}_p$ is the conversion factor from bolometric to differential luminosities, and $\xi_{\rm CR}$ is the CR loading parameter defined against the x-ray luminosity, where $P_{\rm CR}/P_{\rm th}\sim0.01$ corresponds to $\xi_{\rm CR}\sim0.1$ in our model. 
The ENB and EGB are dominated by AGN with $L_{X}\sim{10}^{44}~{\rm erg}~{\rm s}^{-1}$~\cite{Ueda:2014tma}, for which the effective local number density is $\rho_X\sim5\times{10}^{-6}~{\rm Mpc}^{-3}$~\cite{Murase:2016gly}.

The $pp$, $p\gamma$ and Bethe-Heitler processes all initiate cascades, whose emission appears in the MeV range. 
Thanks to the dominance of the Bethe-Heitler process, AGN responsible for the medium-energy ENB should contribute a large fraction $\gtrsim10-30$\% of the MeV EGB.  

When turbulent acceleration operates, the reacceleration of secondary pairs populated by cascades~\cite{Murase:2011cx} can naturally enhance the gamma-ray flux. The critical energy of the pairs, $\varepsilon_{e,\rm cl}$, is determined by the balance between the acceleration time $t_{\rm acc}$ and the electron cooling time $t_{\rm e-cool}$ (see Supplemental Material and Refs.~\cite{Murase:2011cx,2011PhRvL.107c5004H}). We find that the condition for the reacceleration is rather sensitive to $B$ and $t_{\rm acc}$. For example, with $\beta=3$ and $q=1.5$, the reaccelerated pairs can upscatter x-ray photons up to $\sim{(\varepsilon_{e,\rm cl}/m_ec^2)}^2\varepsilon_{X}\simeq3.4~{\rm MeV}~{(\varepsilon_{e,\rm cl}/30~{\rm MeV})}^2(\varepsilon_{X}/1~{\rm keV})$, which may lead to the MeV gamma-ray tail.  
This possibility is demonstrated in Fig.~\ref{fig:diffuse}, and the effective number fraction of reaccelerated pairs is constrained as $\lesssim0.1$\%.

\begin{figure}[tb]
\begin{center}
\includegraphics[width=\linewidth]{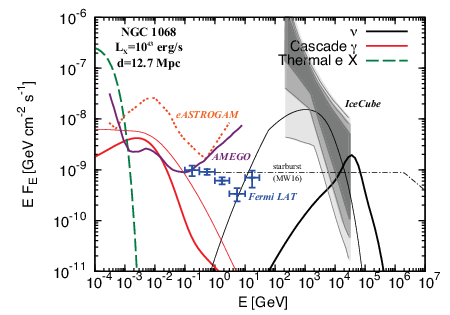}
\caption{Point source fluxes of all flavor neutrinos and gamma rays from a nearby AGN, NGC 1068. 
The ten-year IceCube data~\cite{Aartsen:2019fau} and the {\it Fermi} gamma-ray data~\cite{Lamastra:2016axo} are shown. 
For {\it eASTROGAM}~\cite{DeAngelis:2016slk} and {\it AMEGO}~\cite{Moiseev:2017mxg} sensitivities, the observation time of $10^6$~s is assumed. 
Solid thick (thin) curves are for $\eta=10$ and $P_{\rm CR}/P_{\rm th}=0.7$\% ($\eta=70$ and $P_{\rm CR}/P_{\rm th}=30$\%), respectively. 
For comparison, a neutrino flux in the starburst scenario of Murase and Waxman~\cite{Murase:2016gly} is overlaid. 
}
\label{fig:NGC1068}
\end{center}
\end{figure}

%
{\it Multimessenger tests.---}Our corona model robustly predicts $\sim0.1-10$~MeV gamma-ray emission in either a synchrotron or an inverse Compton cascade scenario, without any primary electron acceleration (see Fig.~\ref{fig:NGC1068}). 
A large flux of 10--100~TeV neutrinos should be accompanied by the injection of Bethe-Heitler pairs in the 100--300~GeV range (see Supplemental Material for details) and form a fast cooling $\varepsilon_e^{-2}$ spectrum down to MeV energies in the steady state. 
In the simple inverse Compton cascade scenario, the cascade spectrum is extended up to a break energy at $\sim1-10$~MeV, above which gamma rays are suppressed by $\gamma\gamma\rightarrow e^+e^-$.   
In reality, both synchrotron and inverse Compton processes can be important. The characteristic energy of synchrotron emission from Bethe-Heitler pairs is $\varepsilon_{\rm syn}^{\rm BH}\sim1~{\rm MeV}~{B}_{2.5}{(\varepsilon_p/0.5~{\rm PeV})}^2$~\cite{Murase:2010va}. Because disk photons lie in the $\sim1-10$~eV range, the Klein-Nishina effect is important for the Bethe-Heitler pairs. 
Synchrotron cascades occur if the photon energy density is smaller than $\sim10B^2/(8\pi)$, i.e., $B\gtrsim170~{\rm G}~L_{\rm disk,45.3}^{1/2}{(\mathcal R/30)}^{-1}R_{S,13.5}^{-1}$. 

The detectability of nearby Seyferts such as NGC 1068 and ESO 138-G001 is crucial for testing the model. MeV gamma-ray detection is promising with future telescopes like {\it eASTROGAM}~\cite{DeAngelis:2016slk}, {\it GRAMS}~\cite{Aramaki:2019bpi}, and {\it AMEGO}~\cite{Moiseev:2017mxg}, e.g., {\it AMEGO}'s differential sensitivity suggests that point sources with $L_X\sim10^{44}~{\rm erg}~{\rm s}^{-1}$ are detectable up to $d\sim70-150$~Mpc. At least a few of the brightest sources will be detected, and detections or nondetections of the MeV gamma-ray counterparts will support or falsify our corona model as the origin of $\sim30$~TeV neutrinos.
Interestingly, as demonstrated in Fig.~\ref{fig:NGC1068}, our corona model can explain the $\sim3\sigma$ excess neutrino flux from NGC 1068~\cite{Aartsen:2019fau}. It also predicts that the x-ray brightest Seyferts (that are more in the southern sky) can be detected as neutrino point sources by IceCube-Gen2 and KM3Net (see also Supplemental Material, which includes Refs.~\cite{Aartsen:2018ywr,Aartsen:2014njl,Adrian-Martinez:2016fdl}).

%
{\it Summary and discussion.---}We presented a new AGN corona model that can explain the medium-energy neutrino data. 
The observed disk-corona SEDs and known empirical relations enabled us to estimate model parameters, with which we solved particle transport equations and consistently computed subsequent electromagnetic cascades.  
Our coronal emission model provides clear, testable predictions relying on the neutrino--gamma-ray relationship that are largely independent of CR spectra. In particular, the dominance of the Bethe-Heitler pair production process is a direct consequence of the observed SEDs, leading to a natural connection with MeV gamma rays.  
Nearby Seyferts such as NGC 1068 and ESO 138-G001 will be promising targets for future MeV gamma-ray telescopes such as {\it eASTROGAM} and {\it AMEGO}. 
A good fraction of the MeV EGB may come from RQ AGN especially with secondary reacceleration, in which gamma-ray anisotropy searches should also be powerful~\cite{Inoue:2013vza}. 
Neutrino multiplet~\cite{Murase:2016gly} and stacking searches with x-ray bright AGN are also promising, as encouraged by the latest neutrino source searches~\cite{Aartsen:2019fau}. 
 
The proposed tests are crucial for unveiling nonthermal phenomena in the vicinity of SMBHs. 
In Seyferts and quasars, the plasma density is so high that a vacuum polar gap is diminished and GeV--TeV gamma rays are blocked. 
MeV gamma rays and neutrinos can escape and serve as a smoking gun of hidden CR acceleration that cannot be probed by x rays and lower-energy photons. Our results also strengthen the importance of further theoretical studies of disk-corona systems. Simulations on turbulent acceleration in coronae and particle-in-cell computations of magnetic reconnections are encouraged to understand the CR acceleration in such systems. Global MHD simulations will also be relevant to examine other speculative postulates such as accretion shocks~\cite{1986ApJ...304..178K,Stecker:1991vm,Szabo:1994qx,Stecker:1995th} or colliding blobs~\cite{AlvarezMuniz:2004uz}, and to reveal the origin of low-frequency emission~\cite{Inoue:2014bwa,Inoue:2018kbv}.  

\vspace{0.3cm}
\acknowledgments
We thank Francis Halzen and Ali Kheirandish for discussion about NGC 1068 (in April 2019). 
K.M. acknowledges the invitation to the {\it AMEGO} Splinter meeting at the 233rd AAS meeting (held in January 2019), where preliminary results were presented. 
This work is supported by Alfred P. Sloan Foundation, NSF Grants No.~PHY-1620777 and No.~AST-1908689, and JSPS KAKENHI No.~20H01901 (K.M.), JSPS Oversea Research Fellowship, IGC Fellowship, JSPS Research Fellowship, and JSPS KAKENHI No.~19J00198  (S.S.K.), NASA NNX13AH50G, and Eberly Foundation (P.M.).

{\it Note added.---} While we were finalizing this project, we became aware of arXiv:1904.00554~\cite{Inoue:2019fil}. We thank Yoshiyuki Inoue for prior discussions. 
Both works are independent and complementary, but there are notable differences. 
First, we consistently calculated cosmic-ray and secondary neutrino and gamma-ray spectra based on the standard picture of magnetized coronae, rather than by hypothesized ``free-fall'' accretion shocks. The former picture is supported by recent simulations of the magnetorotational instability. 
Second, we focused on the mysterious origin of 10--100~TeV neutrinos (see Ref.~\cite{Murase:2015xka} for general arguments), for which the Bethe-Heitler suppression is relevant. The pileup and steep tail in cosmic-ray spectra due to their cooling are also considered. 
Third, we computed electromagnetic cascades, which is essential to test the scenario for IceCube neutrinos. Our model does not need the assumption on primary electrons.

\bibliography{kmurase,sskprl}

\pagebreak
\setcounter{equation}{0}
\setcounter{figure}{0}
\setcounter{table}{0}
\setcounter{section}{0}
\setcounter{page}{1}
\makeatletter
\renewcommand{\theequation}{S\arabic{equation}}
\renewcommand{\thefigure}{S\arabic{figure}}
\renewcommand{\thetable}{S\arabic{table}}
\newcommand\ptwiddle[1]{\mathord{\mathop{#1}\limits^{\scriptscriptstyle(\sim)}}}


\section{Supplemental Material}
\subsection{Disk-corona spectra and properties of coronae}
\begin{table}[b]
\caption{Physical quantities deduced in our model (for ${\mathcal R}=30$) as a function of $L_X$. Units are [erg $\rm s^{-1}$] for $L_X$ and $L_{\rm disk}$, [$\msun$] for $M$, and [$\rm cm^{-3}$] for $n_p$. Note that $\theta_e\equiv k_BT_e/(m_ec^2)$ is the normalized electron temperature.
\label{tab:quantities}
}
\begin{ruledtabular}
\scalebox{1.0}{
\begin{tabular}{c||cc|ccc|c}
 $\log L_X$ & $\log L_{\rm disk}$ & $\log M$ & $\Gamma_X$ & $\theta_e$ &  $\tau_T$ & $\log n_p$\\
 \hline
42.0 & 43.0 & 6.51 & 1.72 & 0.27 & 0.59 & 10.73\\
43.0 & 44.2 & 7.25 & 1.80 & 0.23 & 0.52 & 9.93\\
44.0 & 45.4 & 8.00 & 1.88 & 0.20 & 0.46 & 9.13\\
45.0 & 46.6 & 8.75 & 1.96 & 0.16 & 0.41 & 8.33\\
46.0 & 47.9 & 9.49 & 2.06 & 0.12 & 0.36 & 7.53\\
\end{tabular}
}
\end{ruledtabular}
\end{table}

Here we describe details on the modeling of disk-corona spectral energy distributions (SEDs) that are constructed from observational data and empirical relationships. 

For a disk component, we use the averaged SEDs in Ref.~\cite{ho08}, which are given as a function of the Eddington ratio, $\lambda_{\rm Edd}=L_{\rm bol}/L_{\rm Edd}$, where $L_{\rm bol}$ and $L_{\rm Edd}$ are bolometric and Eddington luminosities, respectively. 
The disk spectrum is expected to have a cutoff at $\varepsilon_{\rm disk, cut}\approx2.8k_BT_{\rm disk}$, where $T_{\rm disk}\approx0.49(GM\dot{M}/72\pi\sigma_{\rm SB}R_{S}^3)^{1/4}$ is the maximum effective temperature of the disk near the central supermassive black hole (SMBH) (e.g., Ref.~\cite{pri81}). Here, $M$ is the SMBH mass, $\dot M$ is the mass accretion rate, $R_{S}=2GM/c^2$ is the Schwarzschild radius, and $\sigma_{\rm SB}$ is the Stefan-Boltzmann constant. For a standard disk, one may use $\dot M\approx L_{\rm bol}/(\eta_{\rm rad}c^2)=\lambda_{\rm Edd}L_{\rm Edd}/(\eta_{\rm rad}c^2)$ with a radiative efficiency of $\eta_{\rm rad}=0.1$~\cite{KFM08a}. 
Although the SEDs in Ref.~\cite{ho08} extend to low energies, we only consider disk photons with $\varepsilon_{\rm disk}>2$ eV because infrared photons would come from a dust torus. The ultraviolet and X-ray data are interpolated with an exponential cutoff taken into account, which gives conservative estimates on target photon densities for cosmic-ray interactions. 

The coronal spectrum can be modeled by a power law with an exponential cutoff. The photon index, $\Gamma_X$, is correlated with $\lambda_{\rm Edd}$ as $\Gamma_X\approx0.167\times\log(\lambda_{\rm Edd})+2.0$ (see Eq.~2 of Ref.~\cite{2017MNRAS.470..800T}), and the cutoff energy is given by $\varepsilon_{X,\rm cut}\sim[-74\log(\lambda_{\rm Edd})+1.5\times10^2]$~keV~\cite{2018MNRAS.480.1819R}. 
The x-ray luminosity $L_X$ can be converted into $L_{\rm bol}$ following Ref.~\cite{2007ApJ...654..731H}, by which the disk-corona SEDs are constructed as a function of $L_X$ and $M$. 

To evaluate the cumulative neutrino and gamma-ray intensities from all active galactic nuclei (AGN) along the line-of-sight, we convolute the x-ray luminosity function with neutrino and gamma-ray spectra calculated as a function of $L_X$. In this work, instead of using the SMBH mass function, we assume the typical SMBH mass estimated by $M\approx2.0\times10^7~M_\odot~(L_X/1.16\times10^{43}~{\rm erg}~{\rm s}^{-1})^{0.746}$~\cite{2018arXiv180306891M}.
The resulting SMBH mass $M$ and disk luminosity $L_{\rm disk}(\sim0.5L_{\rm bol})$ are given in Table~\ref{tab:quantities}. 

The x-ray cutoff energy is also used to estimate the coronal electron temperature $T_e\approx\varepsilon_{X,\rm cut}/(2k_B)$. Ref.~\cite{2018MNRAS.480.1819R} gave $\tau_T\approx 10^{(2.16-\Gamma_X)/1.06}{(k_BT_e/{\rm keV})}^{-0.3}$ in a slab geometry, by which the Thomson optical depth $\tau_T$ is determined. 
For a given SMBH mass, the coronal radius $R$ is parameterized by the normalized coronal radius $\mathcal R$ (that is set to 30 in this work). Then the nucleon density $n_p$ is determined by $\tau_T$ and ${\mathcal R}$.  
In Table~\ref{tab:quantities}, we summarize $\Gamma_X$, $kT_e$, $\tau_T$ and $n_p$, which are obtained in the model.

\subsection{Timescales for thermal particles}
\begin{figure}[tb]
\begin{center}
\includegraphics[width=1.0\linewidth]{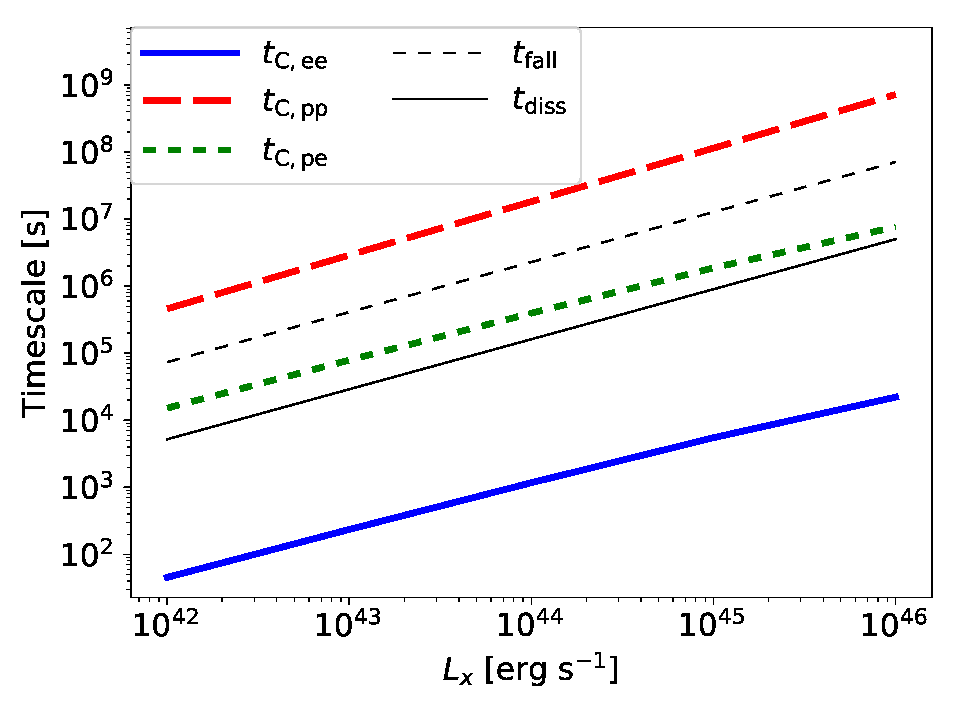}
\caption{Infall, dissipation, and Coulomb relaxation timescales as a function of $L_X$. This shows that the plasma is collisionless for protons, whereas it is collisional for electrons.}
\label{fig:plasma}
\end{center}
\end{figure}

To accelerate particles through stochastic acceleration in turbulence, the Coulomb relaxation timescales at their injection energy should be longer than the dissipation timescale. We discuss these plasma timescales in this subsection. The infall timescale is expected to be similar to that of the advection dominated accretion flow~\cite{ny94,YN14a}: 
\begin{equation}
t_{\rm fall}\approx \frac{{\mathcal R}R_S}{\alpha V_K},
\label{eq:tfall}
\end{equation}
where $\alpha$ is the viscous parameter in the accretion flow~\cite{Shakura:1972te}, ${\mathcal R}$ is the normalized coronal radius, and $R_S$ is the Schwarzschild radius. Assuming that the coronal dissipation is related to some magnetic process like reconnections, the dissipation timescale is expressed as 
\begin{equation}
t_{\rm diss}\approx \frac{H}{V_A} \approx \frac{{\mathcal R}R_S}{V_K}\sqrt{\frac{\beta}{2}},
\end{equation}
which can be short especially for low $\beta$. 
The relaxation timescales for electrons and protons are~\cite{tk85,ktt14}
\begin{equation}
t_{C,ee}\approx \frac{4\sqrt{\pi} \theta_e^{3/2}}{n_p\sigma_Tc\ln\Lambda},
\end{equation}
\begin{equation}
t_{C,pp}\approx \frac{4\sqrt{\pi} \theta_p^{3/2}}{n_p\sigma_Tc\ln\Lambda}\left(\frac{m_p}{m_e}\right)^2 ,
\end{equation}
\begin{equation}
t_{C,pe}\approx \sqrt{\frac{\pi}{2}}\frac{\left(\theta_p+\theta_e\right)^{3/2}}{n_p\sigma_Tc\ln\Lambda}\left(\frac{m_p}{m_e}\right),
\end{equation}
where $\theta_i =k_BT_i/(m_ic^2)$ ($i=p$ or $e$), $\ln\Lambda\sim20$ is the Coulomb logarithm, and we may consider a proton-electron plasma~\cite{2018MNRAS.480.1819R}. 
We plot these timescales as a function of $L_X$ in Fig.~\ref{fig:plasma} for $\beta=1$ and $\alpha=0.1$. We see that among the five timescales $t_{C,ee}$ and $t_{C,pp}$ are the shortest and longest, respectively. This means that electrons are easily thermalized while nonthermal protons are naturally expected and could be accelerated through stochastic acceleration. 
In order to form a two-temperature corona that is often discussed in the literature (e.g., Refs.~\cite{DiMatteo:1997cy,Cao:2008by}), the dissipation timescale should be shorter than the proton-electron relaxation time. Because $t_{\rm diss}\lesssim t_{C,pe}$ is satisfied for the range of our interest, one may expect the two-temperature corona to be formed. Note that electrons in coronae would be heated by magnetic fields from the disk to explain $L_X$ and $kT_e$.  

\subsection{Timescales for high-energy protons}
\begin{figure*}[tb]
\begin{minipage}{0.32\hsize}
\begin{center}
\includegraphics[width=\linewidth]{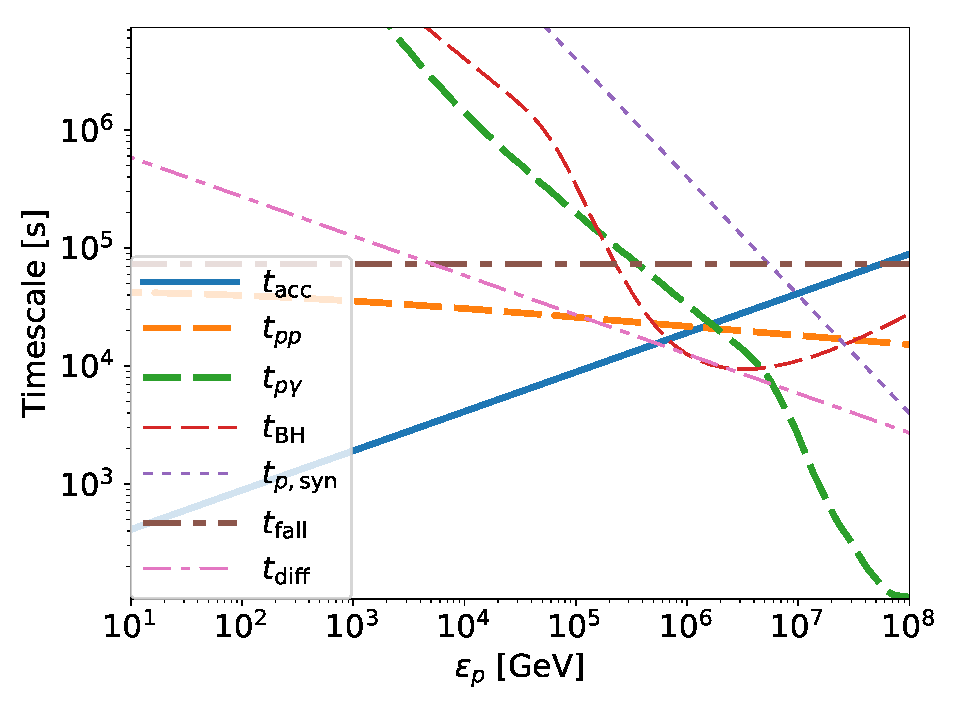}
\end{center}
\end{minipage}
\begin{minipage}{0.32\hsize}
\begin{center}
\includegraphics[width=\linewidth]{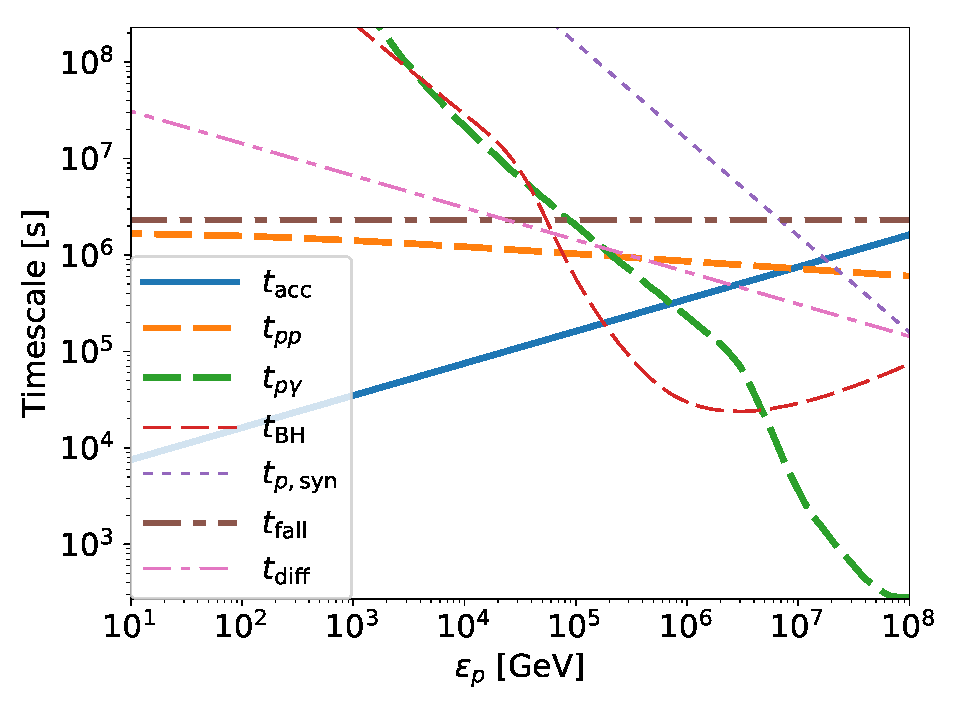}
\end{center}
\end{minipage}
\begin{minipage}{0.32\hsize}
\begin{center}
\includegraphics[width=\linewidth]{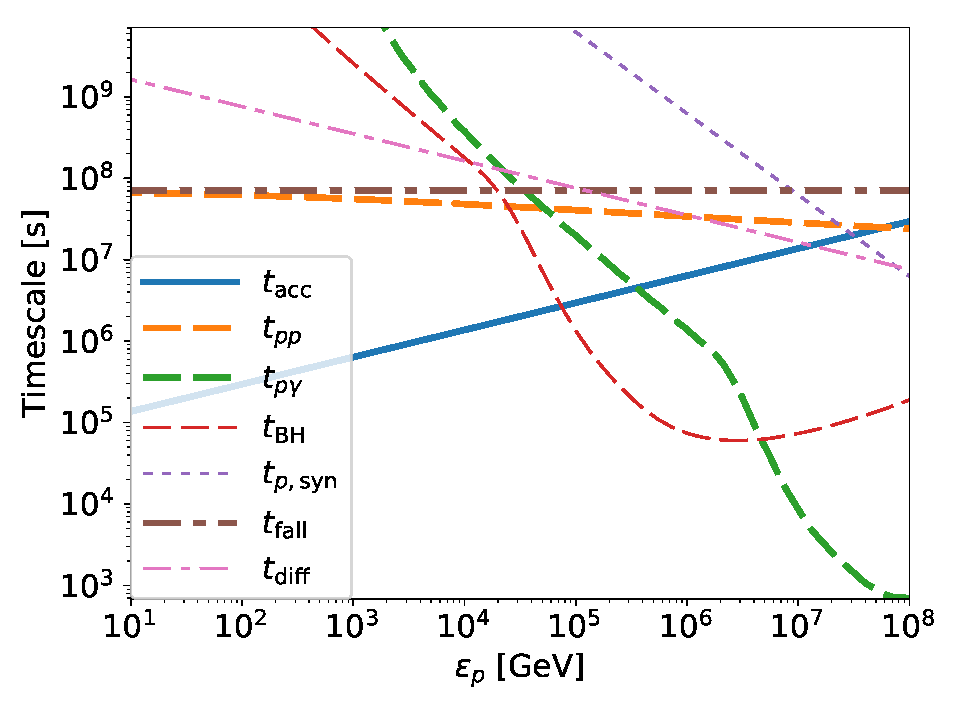}
\end{center}
\end{minipage}
\caption{Comparison of timescales for high-energy protons in the cases with $L_X=10^{42}$ (left), $10^{44}$ (middle), and $10^{46}\rm~erg~s^{-1}$ (right).}
\label{fig:proton_times}
\end{figure*}

Nonthermal proton spectra are determined by the balance among particle acceleration, cooling, and escape processes. We consider stochastic acceleration by turbulence, and take account of infall and diffusion as escape processes. We also treat inelastic $pp$ collisions, photomeson production, Bethe-Heitler pair production, and synchrotron radiation as relevant cooling processes. 

It is well known that stochastic acceleration is modeled as a diffusion phenomenon in momentum or energy space (e.g., Refs.~\cite{Becker:2006nz,Stawarz:2008sp,lyn+14,Kimura:2016fjx,Kimura:2018clk}). 
Assuming gyro-resonant scattering through turbulence with a power spectrum of $P_k\propto k^{-q}$, the acceleration time is written as~\cite{Dermer:1995ju,Murase:2011cx,Dermer:2014vaa,Kimura:2014jba}
\begin{equation}
t_{\rm acc} \approx \eta \left(\frac{c}{V_A}\right)^{2}\frac{H}{c} \left(\frac{\varepsilon_p}{eBH}\right)^{2-q},\label{eq:tacc}
\end{equation}
where $\eta^{-1}=8\pi \int dk\, P_k/B^2$ is the turbulence strength parameter, $V_A=B/\sqrt{4\pi m_p n_p}$ is the Alfv\'en velocity, and $R$ is the effective size of the coronal region.  
The infall time is given by Eq.~(\ref{eq:tfall}). Using the same scattering process for the stochastic acceleration, the diffusive escape time is estimated to be~\cite{Dermer:1995ju,Stawarz:2008sp,Kimura:2014jba}
\begin{equation}
t_{\rm diff}\approx \frac{\eta'}{\eta}\frac{H}{c} \left(\frac{eBH}{\varepsilon_p}\right)^{2-q},
\end{equation}
where $\eta'$ is a prefactor, which we set to $\eta'=9$~\cite{Stawarz:2008sp} within uncertainties of our model with $H\lesssim R$. 
The cooling rate by $pp$ inelastic collisions is given by
\begin{equation}
t_{pp}^{-1} \approx n_p \kappa_{pp}\sigma_{pp} c,
\end{equation}
where $\sigma_{pp}$ and $\kappa_{pp}\approx0.5$ are the cross section and inelasticity for $pp$ interactions, as implemented in Refs.~\cite{Murase:2017pfe,Murase:2018okz}. 
The photomeson production energy loss rate is calculated by
\begin{equation}
t_{p\gamma}^{-1}\approx\frac{c}{2\gamma_p^2}\int_{\overline \varepsilon_{\rm th}}^\infty{d}\overline \varepsilon_\gamma\sigma_{p\gamma}\kappa_{p\gamma}\overline \varepsilon_\gamma\int_{\overline \varepsilon_\gamma/(2\gamma_p)}^\infty{d}\varepsilon_\gamma \varepsilon_\gamma^{-2}\frac{dn_\gamma}{d\varepsilon_\gamma},\label{eq:tpg}
\end{equation}
where $\gamma_p=\varepsilon_p/(m_pc^2)$ is the proton Lorentz factor, $\overline \varepsilon_{\rm{th}}\approx145$~MeV is the threshold energy for the photomeson production, $\overline \varepsilon_\gamma$ is the photon energy in the proton rest frame, $\sigma_{p\gamma}$ and $\kappa_{p\gamma}$ are the cross section and inelasticity, respectively, and the normalization is given by $\int d\varepsilon_\gamma \varepsilon_\gamma (dn_\gamma/d\varepsilon_\gamma)\approx(1+\tau_T)(L_{\rm disk}+L_X)/(2\pi R^2c)$, where $\mathcal V=(4\pi/3)r^3=2\pi{(\mathcal RR_S)}^2H$ is the volume of the coronal region and $r/c$ is used for the photon and neutrino escape time. 
We utilize the fitting formula based on GEANT4 for $\sigma_{p\gamma}$ and $\kappa_{p\gamma}$, which are used in Ref.~\cite{Murase:2008mr}. 
The Bethe-Heitler energy loss rate ($t_{\rm BH}^{-1}$) is written in the same form of Eq.~(\ref{eq:tpg}) by replacing the cross section and inelasticity with $\sigma_{\rm BH}$ and $\kappa_{\rm BH}$, respectively, where we use the fitting formula given in Refs.~\cite{1992ApJ...400..181C} and \cite{SG83a}. 
Finally, the synchrotron timescale for protons is given by 
\begin{equation}
 t_{p,\rm syn}\approx \frac{6\pi m_p^4c^3}{m_e^2\sigma_T B^2 \varepsilon_p}.
\end{equation}

We plot the times scales in Fig.~\ref{fig:proton_times} with a parameter set of ${\mathcal R}=30$, $\beta=1$, $\alpha=0.1$, $\eta=10$, and $q=5/3$. We can see that particle acceleration is limited by interactions with photons except for $L_X=10^{42}\rm~erg~s^{-1}$. 
In the lowest-luminosity case, the photomeson production, the Bethe-Heitler process, the $pp$ reaction, and the diffusive escape rates are comparable to the acceleration rate around $10^6$~GeV. In the other cases, the Bethe-Heitler process hinders the acceleration and the maximum energy is reduced to $\sim10^5-10^6$~GeV due to larger photon number densities of more luminous Seyferts.

\subsection{Spectra of nonthermal protons}
%
\begin{figure}[b]
\begin{center}
\includegraphics[width=0.95\linewidth]{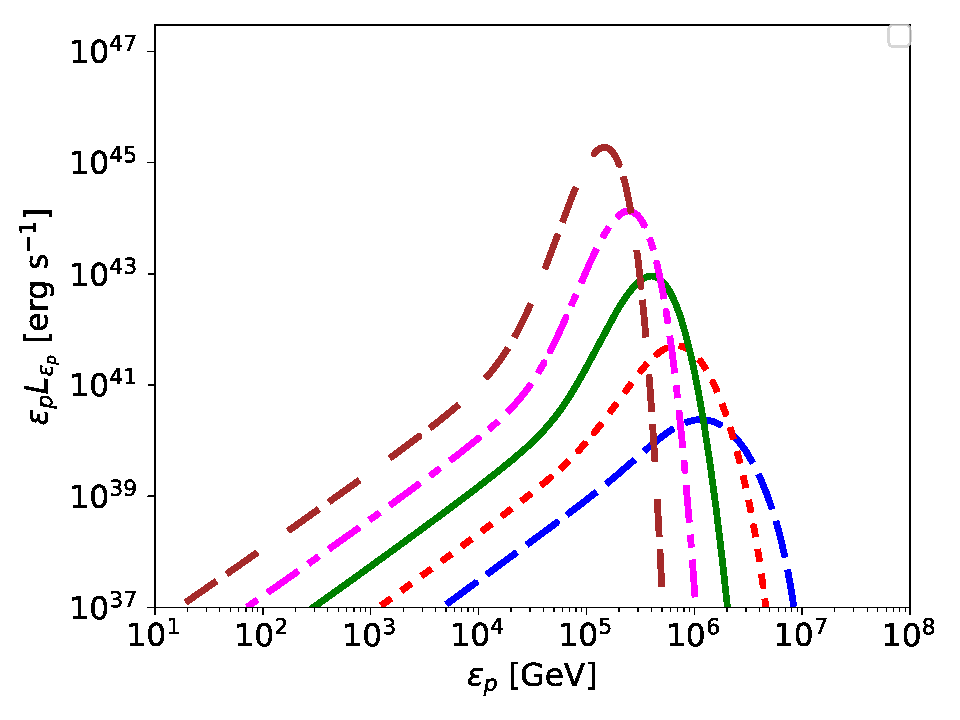}
\caption{CR proton differential luminosities for $L_X=10^{42}$, $10^{43}$, $10^{44}$, $10^{45}$, and $10^{46}\rm~erg~s^{-1}$ (from bottom to top).}
\label{fig:CR}
\end{center}
\end{figure}
\begin{figure*}[tb]
\begin{minipage}{0.32\hsize}
\begin{center}
\includegraphics[width=\linewidth]{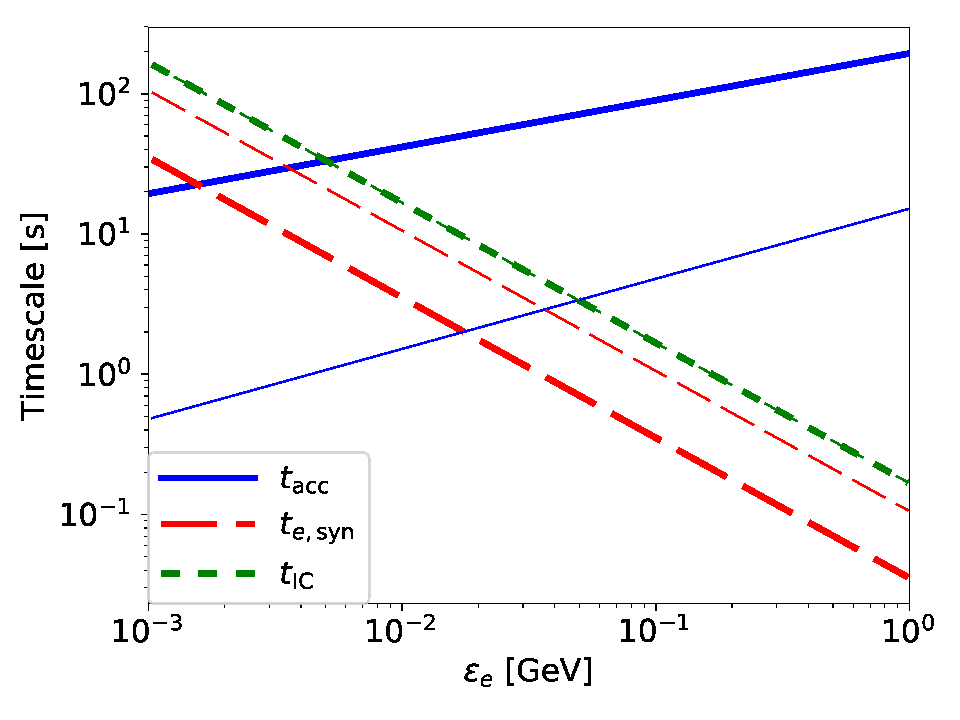}
\end{center}
\end{minipage}
\begin{minipage}{0.32\hsize}
\begin{center}
\includegraphics[width=\linewidth]{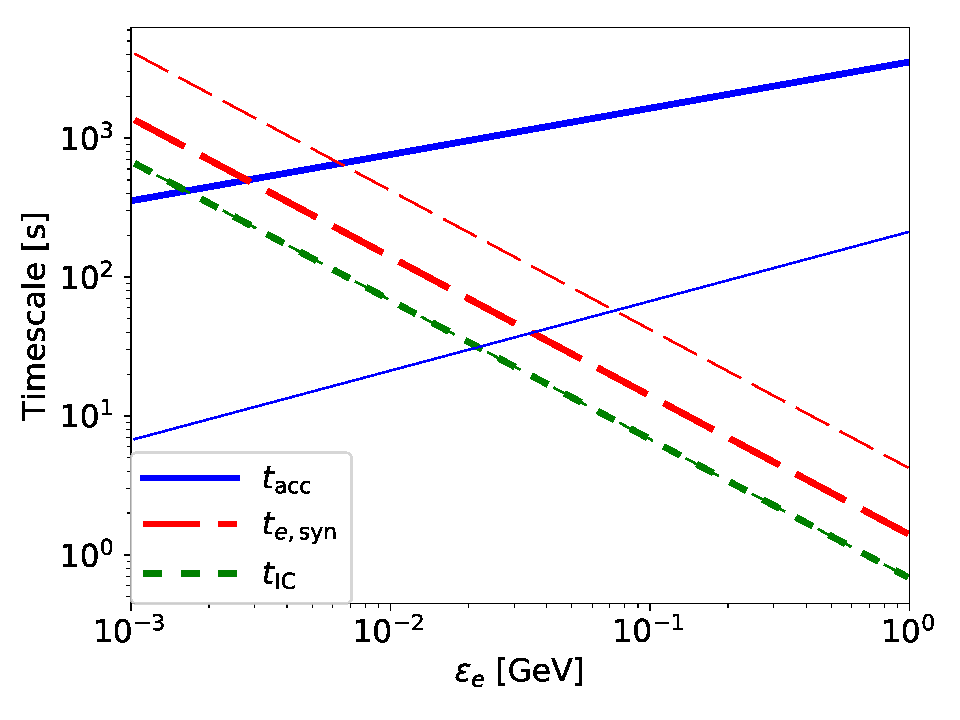}
\end{center}
\end{minipage}
\begin{minipage}{0.32\hsize}
\begin{center}
\includegraphics[width=\linewidth]{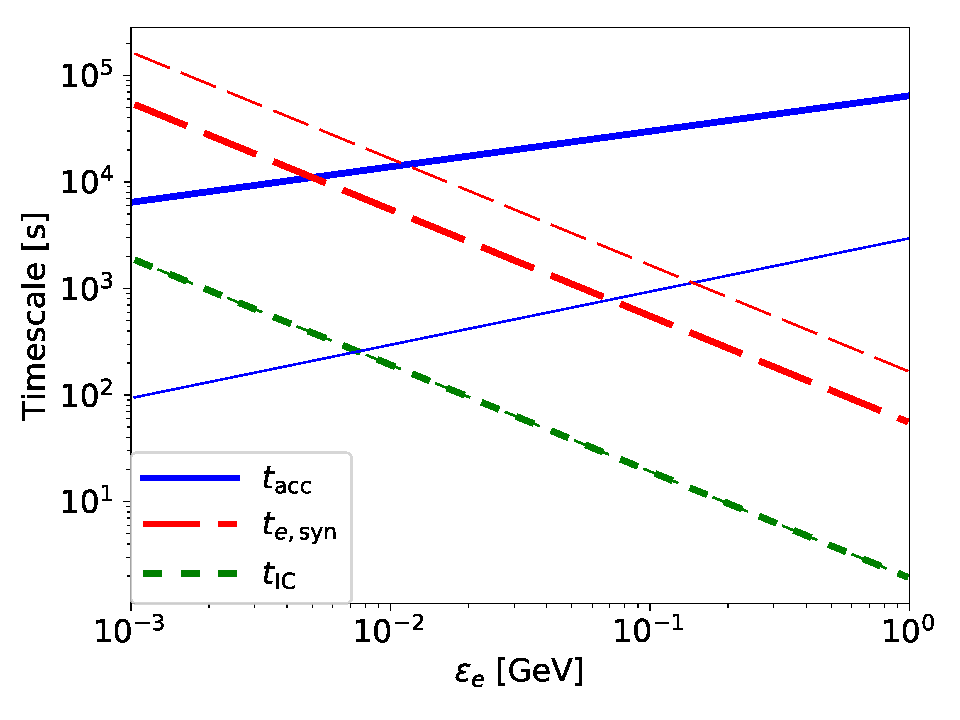}
\end{center}
\end{minipage}
\caption{Comparison of timescales for high-energy electrons and positrons in the cases with $L_X=10^{42}$ (left), $10^{44}$ (middle), and $10^{46}\rm~erg~s^{-1}$ (right).}
\label{fig:pair_times}
\end{figure*}

It is well known that the second order Fermi acceleration can be described by the Fokker-Planck equation (e.g., Refs.~\cite{Becker:2006nz,Stawarz:2008sp,Kimura:2014jba}).  
To obtain spectra of nonthermal cosmic-ray (CR) protons, using the standard Chang-Cooper method~\cite{cc70,1996ApJS..103..255P}, we solve the following equation,
\begin{equation}
\frac{\partial {\mathcal F}_p}{\partial t} = \frac{1}{\varepsilon_p^2}\frac{\partial}{\partial \varepsilon_p}\left(\varepsilon_p^2D_{\varepsilon_p}\frac{\partial {\mathcal F}_p}{\partial \varepsilon_p} + \frac{\varepsilon_p^3}{t_{p-\rm cool}}{\mathcal F}_p\right) -\frac{{\mathcal F}_p}{t_{\rm esc}}+\dot {\mathcal F}_{p,\rm inj},\label{eq:FP}
\end{equation}
where ${\mathcal F}_p$ is the CR distribution function, $D_{\varepsilon_p}\approx \varepsilon_p^2/t_{\rm acc}$ is the diffusion coefficient in energy space, $t_{p-\rm cool}^{-1}=t_{pp}^{-1}+t_{p\gamma}^{-1}+t_{\rm BH}^{-1}+t_{p,\rm syn}^{-1}$ is the total cooling rate, $t_{\rm esc}^{-1}=t_{\rm fall}^{-1}+t_{\rm diff}^{-1}$ is the CR escape rate (for the coronal region), 
and $\dot {\mathcal F}_{p,\rm inj}$ is the injection function that is given by 
\begin{equation}
\dot {\mathcal F}_{p,\rm inj}=\frac{f_{\rm inj}L_X}{4\pi {(\varepsilon_{\rm inj}/c)}^3 {\mathcal V}}\delta(\varepsilon_p-\varepsilon_{\rm inj}),
\end{equation}
where $\varepsilon_{\rm inj}$ is the injection energy and $f_{\rm inj}$ is the injection fraction at $\varepsilon_{\rm inj}$.  
When the normalization is set by the diffuse neutrino flux, we only need $P_{\rm CR}/P_{\rm th}\sim0.1-1$\%. Importantly, this CR energy fraction is only $\sim1-10$\% of the turbulent energy, which is energetically reasonable.
The values of $\varepsilon_{\rm inj}$ and $f_{\rm inj}$ do not affect the resulting spectral shape as long as $\varepsilon_{\rm inj}\ll \varepsilon_{p,\rm cl}$. 
Note that we do not have to specify both of $\varepsilon_{\rm inj}$ and $f_{\rm inj}$, because $f_{\rm inj}$ is larger as $\varepsilon_{\rm inj}$ is higher. 
For $\varepsilon_{\rm inj} < \varepsilon_p \ll \varepsilon_{p,\rm cl}$, it is known that the stochastic acceleration mechanism predicts a very hard spectrum, $dN_{\rm CR}/d\varepsilon_p\propto \varepsilon_p^{1-q}$ (e.g., Refs.~\cite{Becker:2006nz,Stawarz:2008sp,Kimura:2014jba}). 
All details are presented in our accompanying paper~\cite{Kimura:2019yjo}. 

We plot $\varepsilon_p L_{\varepsilon_p}\equiv4\pi (\varepsilon_p^4/c^3) F_p {\mathcal V}(t_{\rm esc}^{-1}+t_{p-\rm cool}^{-1})$ in Fig.~\ref{fig:CR}. 
In Table~\ref{tab:NT-quantities}, we tabulate the critical energy for CR protons, $\varepsilon_{p,\rm cl}$, at which the stochastic acceleration balances with energy and escape loss processes. 
For lower values of $L_X$, the critical energy is higher owing to their lower loss rates. We consider two cases, $q=5/3$ and $q=3/2$, which give similar CR proton spectra.
For $L_X \gtrsim 10^{43}\rm~erg~s^{-1}$, owing to inefficient escape, the accelerated protons pile up around $\varepsilon_p\sim \varepsilon_{p,\rm cl}$, which creates a hardening feature around $\varepsilon_p\sim 0.1\varepsilon_{p,\rm cl}$ and a strong cutoff above $\varepsilon_p\sim 2\varepsilon_{p,\rm cl}$. 
Note that the resulting spectra of $p\gamma$ neutrinos are rather insensitive to such hard CR spectra, because they cannot be harder than the neutrino spectrum from pion decay~\cite{Murase:2015xka}. 
The energy density of CRs is typically lower than that of the thermal protons: $P_{\rm CR}/P_{\rm th}$ is of the order of $10^{-2}$ for all the cases that allow us to explain the diffuse neutrino flux (see Table~\ref{tab:NT-quantities}), where $P_{\rm CR}=\int\,d\varepsilon_p (dn_{\rm CR}/d\varepsilon_p)\varepsilon_p/3$ and $P_{\rm th}=n_pk_BT_p$. These CRs do not affect the dynamical structure~\cite{ktt14}.

\begin{table}[b]
\caption{Physical quantities related to nonthermal particles for a given $L_X$. The CR pressures accounting for the 10--100~TeV ENB flux are also shown. The reacceleration of pairs is irrelevant in one case with $\beta=1$ and $q=5/3$, whereas it can occur in the other case with $\beta=3$ and $q=3/2$.
\label{tab:NT-quantities}}
\begin{ruledtabular}
no secondary reacceleration
\scalebox{1.0}{
\begin{tabular}{c||cc|ccc}
$\log L_X$ & $\log t_{\rm fall}$ & $B$ & $P_{\rm CR}/P_{\rm th}$ & $\varepsilon_{p,\rm cl}$ & $\varepsilon_{e,\rm cl}$ \\
 $[\rm erg~s^{-1}]$ &  [s]  & [kG] & [\%] & [TeV] & [MeV] \\
\hline
42.0 & 4.87 & 3.4 & 0.32 & 300 & $<5$  \\
43.0 & 5.61 & 1.3 & 0.70 & 220 &  $<5$ \\
44.0 & 6.36 & 0.53 & 1.3 & 140 & $<5$  \\
45.0 & 7.11 & 0.21 & 2.2 & 100 & $<5$ \\
46.0 & 7.85 & 0.085 & 3.6 & 70 & $<5$ \\
\end{tabular}
}
allowed secondary reacceleration
\scalebox{1.0}{
\begin{tabular}{c||cc|ccc}
$\log L_X$ & $\log t_{\rm fall}$ & $B$ & $P_{\rm CR}/P_{\rm th}$ & $\varepsilon_{p,\rm cl}$ & $\varepsilon_{e,\rm cl}$\\
 $[\rm erg~s^{-1}]$ & [s] & [kG] & [\%] & [TeV] & [MeV] \\
\hline
42.0 & 4.87 & 1.9 & 0.25& 450 & 27 \\
43.0 & 5.61 & 0.77 & 0.48 & 320 & 25 \\
44.0 & 6.36 & 0.31 & 0.76 & 230 & 20 \\
45.0 & 7.11 & 0.12 & 1.2 & 160 & 13 \\
46.0 & 7.85 & 0.049 & 1.6 & 110 & 8 \\
\end{tabular}
}
\end{ruledtabular}
\end{table}

\subsection{Timescales for high-energy pairs}
Even if primary electrons are not accelerated through the turbulence due to their efficient Coulomb losses, sufficiently high-energy electron-positron pairs that are injected via hadronic processes and populated through electromagnetic cascades can be accelerated to higher energies without suffering from Coulomb losses, as studied in the literature of gamma-ray bursts~\cite{Murase:2011cx}. 
In the meantime, such high-energy pairs also cool down by synchrotron radiation and inverse Compton emission. In Seyferts, the energy of the dominant target photons for inverse Compton scatterings is around 1--10~eV, implying that the Klein-Nishina effect is unimportant at $\varepsilon_e\lesssim100-300$~GeV. 
If the secondary pairs can be accelerated by the turbulence, the acceleration and cooling processes are balanced at the critical energy, $\varepsilon_{e,\rm cl}$, and this effect is relevant if $\varepsilon_{e,\rm cl}$ is higher than the energy of the thermal protons. The timescale of the turbulent reacceleration is given by Eq.~(\ref{eq:tacc}) above the thermal energy of protons, $k_BT_p\simeq5.2~{({\mathcal R}/30)}^{-1}$ MeV. 
Below this energy, the turbulent power spectrum should become steeper due to kinetic effects~\cite{2011PhRvL.107c5004H}, and the reacceleration time is considerably longer than that by Eq.~(\ref{eq:tacc}). 

The electron synchrotron cooling timescale is 
\begin{equation}
t_{e,\rm syn}\approx \frac{6\pi m_e c}{\sigma_T B^2 \gamma_e},
\end{equation}
where $\gamma_e=\varepsilon_e/(m_ec^2)$ is the electron Lorentz factor.
Then, the electron inverse Compton cooling time in the Thomson limit is estimated to be 
\begin{equation}
t_{\rm IC}\approx \frac{3 m_e c}{4 \sigma_T U_\gamma \gamma_e} ,
\end{equation}
where $U_\gamma$ is the target photon energy density. 
Note that the Klein-Nishina effect is accurately taken into account in the calculations of cascaded photon spectra~\cite{Murase:2017pfe,Murase:2018okz,Kimura:2019yjo}. 

We plot timescales for high-energy pairs in Fig.~\ref{fig:pair_times}, where thin (thick) lines are for the case with reacceleration allowed (forbidden), respectively. 
In the case of no secondary reacceleration, we have $\varepsilon_{e,\rm cl}\ll 5$~MeV, which is the critical energy is lower than the thermal energy of protons. 
The waves are expected to be dissipated at the relevant scales, and the stochastic acceleration of pairs is unlikely in all the range of $L_X$. 
In the other case allowing reacceleration, electrons and positrons injected via hadronic cascades can be maintained with energies around $\sim10-30$~MeV, depending on $L_X$ (see Table~\ref{tab:NT-quantities}), and the reaccelerated pairs can naturally enhance the MeV gamma-ray emission. 
To see the effect of the reacceleration, we take the approximate approach, in which we calculate electromagnetic cascades assuming that the pairs do not cool down below the critical energy that is determined by the balance between the acceleration time and cooling time. 
For $L_X=10^{42}\rm~erg~s^{-1}$, the synchrotron cooling is more likely to be important, while for $L_X\gtrsim 10^{43}\rm~erg~s^{-1}$, the inverse Compton cooling is dominant. The critical electron energy, $\varepsilon_{e,\rm cl}$, at which the cooling and reacceleration balance with each other is lower for a higher value of $L_X$ due to the more efficient inverse Compton cooling. Only when the critical energy is higher than the thermal proton temperature, the secondary reacceleration can occur.
As shown in the main text, the effective number fraction of the reaccelerated pairs is constrained to be less than $\sim0.1$\% not to overshoot the MeV extragalactic gamma-ray background (EGB). This can be consistent with a prediction of stochastic acceleration (see, e.g., Ref.~\cite{Stawarz:2008sp}). If pairs above $\varepsilon_{e,\rm reacc}\sim{\rm a~few}$~MeV can be reaccelerated, the number of pairs in the steady state is reduced by $\sim t_{e-\rm cool}(\varepsilon_{e,\rm reacc})/t_{\rm fall}$ (e.g., $\sim0.05$\% for $L_X=10^{43}\rm~erg~s^{-1}$) compared to the total number of injected pairs during $t_{\rm fall}$. The reacceleration is more difficult in luminous AGN due to small values of $t_{\rm IC}$.

\subsection{Extragalactic neutrino background}
The extragalactic neutrino background (ENB) flux (mean neutrino intensity) is calculated by (e.g., Ref.~\cite{Murase:2014foa})
\begin{eqnarray}
\Phi_\nu&=&\frac{c}{4\pi H_0}\int^{\rm z_{\rm max}} \!\!\! dz \, \frac{1}{\sqrt{{(1+z)}^3\Omega_m+\Omega_\Lambda}}\nonumber\\
&\times& \int \!\!\! dL_X \, \frac{d \rho_X}{dL_X}(z) \frac{L_{E'_\nu}(L_X)}{E'_\nu},
\end{eqnarray}
where $d \rho_X/dL_X$ is the x-ray luminosity function of AGN per comoving volume per luminosity, $L_{E'_\nu}(L_X)$ is the neutrino luminosity per energy, and $z_{\rm max}$ is the maximum value of redshift $z$. Also, $H_0$ is the local Hubble constant, whereas $\Omega_m=0.3$ and $\Omega_\Lambda=0.7$ are cosmological parameters. 
The differential neutrino luminosity, including spectral information, is calculated as a function of the x-ray luminosity, following our model described in this work. 
The x-ray luminosity function of AGN can be described by a smoothly connected double power-law model, and we use Eq.~(14) and Table~4 of Ref.~\cite{Ueda:2014tma}.  
The luminosity function evolves as redshift, for which Eqs.~(16)--(19) and Table~4 of Ref.~\cite{Ueda:2014tma} are adopted. 
It is known that the x-ray luminosity density peaks around a redshift of $z\sim1-2$ with a fast evolution. With the typical redshift evolution of x-ray AGN, the dimensionless parameter $\xi_z$, which represents the redshift evolution of the sources, gives $\xi_z\sim3$~\cite{Murase:2016gly,Waxman:1998yy}.   

The results for the ENB flux from AGN coronae are shown in Fig.~3 in the main text. We consider the sum of all three neutrino flavors.
Note that these AGN are dominated by radio quiet (RQ) AGN that are mostly Seyfert galaxies with a standard accretion disk. 
On the other hand, some AGN are radio loud (RL), which are characterized by the radio loudness and attributed to jet activities. AGN jets are known to be powerful CR accelerators, and blazars and their misaligned counterparts -- radio galaxies -- account for the dominant fraction of the EGB in the 0.1--100~GeV range (e.g., Refs.~\cite{Ajello:2009ip,Ajello:2015mfa}). 
Although the blazar contribution is likely to be subdominant in the neutrino sky~\cite{Murase:2016gly}, RL AGN are among the most promising candidates as the sources of ultrahigh-energy cosmic rays (UHECRs). 
The CRs escaping from the jets can be confined in magnetized environments such as galaxy clusters and groups, generating high-energy neutrinos and gamma rays~\cite{Murase:2008yt,Kotera:2009ms}. 
It is suggested that the jetted AGN embedded in clusters and groups give a ``grand unified'' explanation for UHECRs, sub-PeV neutrinos, and the nonblazar component of the EGB~\cite{Murase:2016gly,Fang:2017zjf}.  
Fig.~\ref{fig:diffuse} demonstrates that total contributions from AGN coronae, jetted AGN, and cosmogenic emission can explain the EGB from MeV to TeV energies as well as the ENB from TeV to PeV energies.
The EGB from blazars~\cite{Ajello:2015mfa} and radio galaxies~\cite{Inoue:2011bm}, and the ENB from galaxy clusters and groups with jetted AGN~\cite{Fang:2017zjf} are shown in Fig.~\ref{fig:diffuse}. 
The latest IceCube data suggest a soft spectrum below 100~TeV with a large ENB flux~\cite{Aartsen:2020aqd}, whereas the higher-energy data above 100~TeV can be described by a harder spectrum with $s\sim2.2-2.3$~\cite{ICRCtrack}. 
Although the current IceCube data do not exclude the single component scenario, they intriguingly hint at a structure in the neutrino spectrum~\cite{Aartsen:2020aqd}, which is compatible with two- or multi-component models~\cite{Chen:2014gxa,Palladino:2018evm}. Indeed, the x-ray luminosity density of AGN is comparable to the CR luminosity of jetted AGN, which are $\sim3\times{10}^{46}~{\rm erg}~{\rm Mpc}^{-3}~{\rm yr}^{-1}$~\cite{Murase:2018utn}, and it is reasonable that the AGN coronae and the jetted AGN embedded in gaseous environments give comparable contributions to the ENB flux with different spectral indices. In addition, it was shown that the 10--100~TeV neutrino data can be explained only by hidden CR accelerators~\cite{Murase:2015xka}, for which our AGN corona model provides a viable explanation and has reasonable energetics consistent with both theory and observations.  

\begin{figure}[t]
\begin{center}
\includegraphics[width=\linewidth]{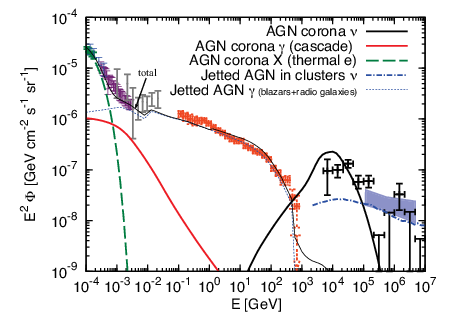}
\caption{A possible unified picture of the EGB and ENB, in which the MeV-TeV EGB and TeV-PeV ENB are simultaneously explained by AGN cores and jetted AGN. The data are taken from {\it Swift} BAT~\cite{Ajello:2008xb} (green), Nagoya balloon~\cite{1975Natur.254..398F} (blue), SMM~\cite{1997AIPC..410.1223W} (purple), COMPTEL~\cite{2000AIPC..510..467W} (gray), {\it Fermi} LAT~\cite{Ackermann:2014usa} (orange), and IceCube for shower (black)~\cite{Aartsen:2020aqd} and upgoing muon track (blue shaded)~\cite{ICRCtrack} events.}
\label{fig:diffuse}
\end{center}
\end{figure}

\subsection{Detectability of the most promising sources}
\begin{figure}[tb]
\begin{center}
\includegraphics[width=\linewidth]{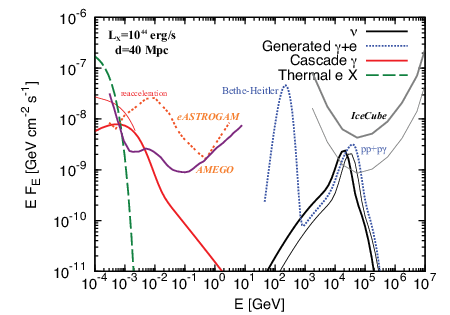}
\caption{Point source fluxes of all flavor neutrinos and gamma rays from a nearby AGN in our corona model. For {\it eASTROGAM}~\cite{DeAngelis:2016slk} and {\it AMEGO}~\cite{Moiseev:2017mxg} sensitivities, the observation time of $10^6$~s is assumed. The IceCube eight-year sensitivity~\cite{Aartsen:2018ywr} and the 5 times better case (cf. Refs.~\cite{Aartsen:2014njl,Adrian-Martinez:2016fdl}) are shown (assuming a declination angle of $\delta=30^\circ$).  
Solid thick (thin) curves are for $\beta=1$ and $q=5/3$ ($\beta=3$ and $q=3/2$ with the effective number fraction of reacceleration 0.05\%), respectively. The CR pressures are given in Table~\ref{tab:NT-quantities}.
}
\label{fig:psflux}
\end{center}
\end{figure}

Our AGN corona model has a prediction for the relationship among the neutrino luminosity, soft-gamma-ray luminosity, and x-ray luminosity. AGN with higher ``intrinsic'' luminosities are better as the sources of high-energy neutrinos, and our model can be uniquely tested by future MeV gamma-ray satellites as well as neutrino experiments such as IceCube-Gen2 and KM3Net. 
In Fig.~\ref{fig:psflux}, we show the case for $L_X=10^{44}~{\rm erg}~{\rm s}^{-1}$ at $d=40$~Mpc (cf. ESO 138-G001). As one can see, the Bethe-Heitler process is crucial, and cascade gamma rays appear in the MeV range via electromagnetic cascades. 

Stacking and cross correlation studies are also powerful, for which the x-ray catalogues, e.g., one from the {\it Swift} BAT AGN Spectroscopic Survey (BASS), should be useful~\cite{BASS}. 
Among Seyfert galaxies listed in BASS, based on the brightest x-ray objects in the 2--10~keV band, we find that the seven brightest targets are Circinus Galaxy, ESO 138-G001, NGC 7582, Cen A, NGC 1068, NGC 424, and CGCG 164-019. In particular, NGC 1068 and CGCG 164-019 are the brightest for upgoing muon neutrino observations by IceCube and IceCube-Gen2. On the other hand, the other Seyferts should be more interesting targets for KM3Net because they are located in the southern sky. 
Note that the merging galaxy NGC 6240 is also an interesting object, but the x-ray emission may not come from the AGN corona.


\end{document}